\documentstyle[12pt]{article}
\def\fnote#1#2{\begingroup\def\thefootnote{#1}\footnote{#2}
    \addtocounter{footnote}{-1}\endgroup}
\def\email{mabe@th.phys.titech.ac.jp}
\def\sato{sato@th.phys.titech.ac.jp}
\def\IP{\relax{\rm I\kern-.18em P}}
\def\CP{\relax {\bf \rm C}{\rm I\kern-.18em P}}
\def\WCP{\relax{\bf \rm WC}{\rm I\kern-.18em P}}
\def\R{\relax{\rm I\kern-.18em R}}
\def\K3{\relax{\rm K3}}
\def\CY{\relax{\rm CY}}

\begin{document}
\pagestyle{empty}
\begin{flushright}
               TIT/HEP-411

\end{flushright}
\vspace{18pt}

\begin{center}
{ \bf  Puzzles  on the Duality  between   \\
\vspace{8pt}   Heterotic  and  Type IIA  Strings
} \\ 
\vspace{4pt}
\vspace{16pt}
Mitsuko Abe \fnote{*}{\email}~
and ~Masamichi Sato\fnote{\dag}{\sato} \\
\vspace{16pt}

{\sl Department of Physics\\
Tokyo Institute of Technology \\
Oh-okayama, Meguro, Tokyo 152, Japan}
\vspace{12pt}

{\bf ABSTRACT}
\vspace{12pt}

\begin{minipage}{4.8in}
      We discuss the possibility of the 
extension of the duality  between  the webs of heterotic 
string and the type IIA string 
to  Calabi-Yau three-folds     
with another K3 fiber by  comparing   the    
dual polyhedron of Calabi-Yau three-folds  
 given by Candelas, Perevalov and Rajesh.

\end{minipage}
\end{center}
\newpage
\pagestyle{plain}
\setcounter{page}{1}


\section{Introduction}

Some  Calabi-Yau three-folds(=CY 3-folds) with base, \CP$^1(1,s)$ and fiber, 
\newline
K3$= \CP^3(u_1, u_2, u_3, u_4)[d]$ 
 are represented in hypersurface in weighted projective 
4-space, \CP$^4$. 
\begin{equation}
\CY3 {\rm -fold} =
 \CP^4(u_1, s u_1, (s+1)u_2, (s+1)u_3, (s+1)u_4))[(s+1)d],
\end {equation}
\noindent
with $ d= \sum_{i=1}^4 u_i$.
\par

The  type IIA string  on 
CY3-folds which have K3 fiber and T$^2$ fiber with at least one  
section  is dual to  the heterotic string on K3 $\times$ T$^2$ 
as pointed out in~\cite{vafa1,vafa2}.  
Much  research has been  explored on CY3-folds with  
$s=1$ and $u_1$=1 cases and  F$_0$ based case~\cite
{vafa1,vafa2,aldazabal,bershadsky,candelas1,candelas2,candelas3}.  
  $s=1$ and $u_1$=1 cases are given by  the  
extremal transitions
  or by  the conifold transitions
 from F$_0$ based CY3-folds \cite{candelas1,candelas2,candelas3}.
 
Thus far, CY3-folds which have been studied  
  are  constructed by using dual polyhedra.
 Some of them  may be related to the hypersurface representations of  eq. (1)
with $s \geq 2$. However,  
 identification between  them    
has some ambiguities and is  complicated~\cite
{candelas4,candelas0,skarke,avram}. 
Therefore, the relation between these results and the 
 duality between the type IIA string  on CY3-folds with $s \geq 2$ 
 to heterotic string on K3 $\times$ T$^2$ 
is not clear   yet.
There have been three   types of web sequences of 
heterotic - type IIA  string  duality 
from the terminal CY3-fold in A series 
\cite{aldazabal,candelas1,candelas2}.

(iii) in \cite{candelas2} may be the subset of  (i)$^\dag$ in 
\cite{candelas3}, 
though the properties of dual polyhedron 
are slightly  different
 \footnote{ The difference between  
dual polyhedron of  (i)$^\dag $ 
and (iii)  is as follows \cite{candelas1,candelas2,candelas3}.  
The dual polyhedra of case  (i)$^\dag$ have the modified 
 dual polyhedra of  K3 fiber.
For the case  (iii) in A series,  the dual polyhedron of 
K3 fiber    is  not modified.  
The highest point in  the  
additional  points is  represented by the  weights of the 
K3 fiber of the terminal A series in this base.
  One   point such  as  $(0,\ast,\ast,\ast)$   is also  
represented by   the part of the  weight of the terminal K3 fiber. 
The following element in SL(4,{\bf Z}) can 
 transform   these polyhedra  
into the dual polyhedron given by  
   the ref. of  \cite{candelas2}.

\[
\left(
\begin{array}{cccc}
1 & 0&1&2 \\
0&1& 2& 3 \\ 
0&0 & 1 &2\\
0&0 &  -1& -1\\
\end{array}
\right).
\]
 In the base of \cite{candelas2},
the additional points make a line with $x_4=-1$. }.
(i)$^\dag$ means the modified (i) in 
\cite {candelas1} with extra tensor multiplets.

\begin{eqnarray}
&{\rm (i)}& 
 G_2^0=I \mid_{G_1^0, {n_T}^0}  ~{\rm with}~   n^0 = \{ 0,1, 2 \}  
\rightarrow G_2^0=\hat G_2 \mid_ {G_1^0,{n_T}^0}~ {\rm with}~   n^0  \geq 3 
 \rightarrow \cdots,    
 \nonumber \\
& \big\{   {\rm (i)}^\dag & 
 (G_2^0, {n_T}^0) \mid_{G_1^0 }  
\rightarrow (G_2, {n_T}) \mid_ {G_1=G_1^0}
 \rightarrow \cdots ,\big\}  
\nonumber \\ 
&{\rm (ii)}&  G_1^0=I \mid_{G_2^0,n_T^0}~{\rm with}~ n^0=j \rightarrow 
 {G_1^0}=\hat G_1 \mid_ {G_2^0, n_T^0}~ {\rm with}~ n^0=j \rightarrow \cdots, 
 \nonumber \\
& \big\{   {\rm  (ii )}^\dag & (G_1^0,n_T^0)\mid_{G_2^0}  \rightarrow 
( {G_1}, n_T)\mid_ {G_2=G_2^0}   \rightarrow \cdots, \big\}
 \nonumber \\ 
&{\rm (iii)}& 
(G_2^0=I, n_T^0 )\mid _{G_1^0=I}~{\rm with}~ n^0=0  \rightarrow (G_2,n_T)
\mid_{G_1=I} ~
\rightarrow \cdots,
\end{eqnarray}
\noindent
where $\hat G_1$ and $\hat G_2$ are non-Abelian gauge 
symmetries and $ 12 \geq j \geq 0$, see Fig. 1 and Fig. 2.
In this paper, we investigate the  possibility of 
following sequence where we picked up CY3-folds and Hodge numbers 
from the list by Hosono et al. \cite{yau}. 
\begin{equation} 
{\rm (iv)} ~
(G_2^0=I, s=1,n_T^0)\mid_{ G_1^0=I} ~{\rm with}~ n^0=2
 \rightarrow (G_2,s=2,n_T)\mid _{G_1=I}  \rightarrow \cdots
\end{equation}
\begin{eqnarray}
&G_1^0~ ({\rm or}~G_1)&:{\rm  the   ~gauge ~symmetry~with ~charged~ matter,~whose~information}
 \nonumber \\
&~&{\rm ~may~come~ from   ~ the~ bottom~ of~dual~polyhedron,}
~\nabla ~
 \nonumber \\
&~&{\rm of~ appropriate~ K3~ fiber~ for~ the~ series~ (ii)~(or~(ii}^\dag))
~\cite{candelas1},
 \nonumber \\
&G_2^0~({\rm or}~G_2) &:{\rm  the~terminal ~gauge ~symmetry~with~ no~ charged~ matter, ~whose}
 \nonumber \\
& &{\rm ~information~may~come~ from  ~ the~ top~ of~dual~polyhedron,}
 \nonumber \\
&& ~\nabla~ {\rm of~ appropriate~K3~ fiber~ for~ the~ series~ (i)~( or~ (i
}^\dag)) ~\cite{candelas1,skarke}. 
\end{eqnarray}

\noindent
For A series, the non-Abelian instanton numbers 
of the vector bundles on K3  are denoted as 
$(k_1^0,k_2^0);~ k_1^0+k_2^0=24$~
\footnote{
Let $m_i^0=m_i$, $m_{iA}^0=m_{iA}$ and $m_{iB}^0 =m_{iB}~ (i=1,2)$ be Abelian instanton numbers.
For B series, non-Abelian instanton numbers and Abelian instanton numbers 
are denoted as 
$(k_1^0,m_1^0:k_2^0,m_2^0);~ k_1^0+k_2^0=18,~ m_1^0=m_2^0=3$. 
For C series, they are denoted as 
$(k_1^0,m_{1A}^0,m_{2A}^0:k_2^0,m_{1B}^0,m_{2B}^0),
~ k_1^0+k_2^0=14,~ m_{1A}^0=m_{1B}^0=3,~
m_{2A}^0=m_{2B}^0=2$ .}.
\noindent
The integer of  $n^0$ is defined by  $k_1^0=12+n^0$ and  
~$k_2^0=12-n^0$~
\footnote{
 The number of non-Abelian instantons should 
be greater than three except zero.  
The terminal group should change  to avoid 
 this situation, which causes  a correction or a modification 
  \cite{aldazabal,candelas2,candelas3}.  
For C series, the non-Abelian and Abelian instanton numbers 
are modified as follows \cite{aldazabal}. 
 In $n^0=5$ case, $ (k_1^0,m_{1A}^0,m_{1B}^0~;~
k_2^0,m_{2A}^0,m_{2B}^0)=(12,3,2;0,3,3)$. 
In $n^0=6$ case, $(k_1^0,m_{1A}^0,m_{1B}^0~;~
k_2^0,m_{2A}^0,m_{2B}^0)=(13,3,2;0,3,3)$.}.
The suffix 0 denotes  terminal case with 
$ n_T=1 $ in each series, where 
  ``terminal case''  means $G_1^0=I$ or  $G_1=I$.
\noindent 
When $\Delta n_T$  
E$_8$  small instantons shrink, in 
the terminal gauge symmetry side,  the 
 non-Abelian instanton numbers   should  be modified as follows 
\cite{aldazabal,candelas3}:
$(k_1,k_2)$,~$ k_1+k_2+\Delta n_T=24$,~$k_1=k_1^0 = 12$,~
$k_2=k_2^0-\Delta n_T=12-
\Delta n_T$
\footnote
{Similarly to A, for B and C, the modifications may  follow 
\newline
\noindent
B-chain~: 
$k_1 + m_1=k_1^0 +m_1^0 = 12$,~
$k_2+m_2 =12 -\Delta n_T  $,
\newline
C-chain ~: $ k_1+m_{1A}+m_{1B} = k_1^0+m_{1A}^0+m_{1B}^0 =
12 $,~$k_2+m_{2A}+m_{2B} =12 -\Delta n_T $, 
\newline
which we will discuss in this article.}.

The additional $\Delta n_T$  numbers of tensor multiplets 
are created and  29 $\Delta n_T$ appears in the anomaly
cancellation condition 
\footnote{
29 is the dual coxetor number of E$_8 - $one \cite{aldazabal,candelas3}. 
The subtraction of one   means the freedom of  fixing the 
place where a tensor multiplet 
is created.
 The dual coxetor number of E$_7 -$one = 17 works for B-chain case 
when a tensor multiplet is created and that of E$_6 -$one =11 for 
the C-chain case \cite{aldazabal,candelas3}.}.
$n_T=n_T^0+\Delta n_T$ with $ n_T^0=1$ is 
the number of the tensor multiplets in D=6 N=1 compactification
\footnote{In D=4 and N=2 case, these  tensor multiplets become 
 vector multiplets.   }. 

\vspace{8pt}
The puzzles  which we would like to  consider       
are as follows:
\newline

I:~Why (iii) and (iv) have the same Hodge numbers for A series ? 
Do (iv) satisfy the duality ?  Though  K3 fibers of CY3-folds in (iv) 
are different from those in (iii),  is there  
any gauge enhancement for  (iv)  ?
\newline 

II:~Is it possible to obtain   
 B or C versions of (iii)  which satisfy the duality 
by means of conifold transitions ? 
 Already,   the method to 
construct dual polyhedron was given by \cite{candelas2,candelas3} 
~about this question.  
They pointed out that their Hodge numbers are obtained by shrinking 
of tensor multiplets \cite{candelas3}.
\newline

III:~Is it possible to obtain  
 B or C versions of (iv) which satisfy the duality by means of conifold 
 transitions ? 
\newline

IV:~ Do the Hodge numbers in B or C series  of (iii)
 coincide with those of (iv)?
\newline

V:~ Are there any 
 hypersurface representing  CY3-folds with $s \geq 2$  
 which   satisfy the duality from $n^0 \geq 4$ ?
\newline

In section 2, we investigate prob. I. 
We also discuss prob. III by using the results  
 of \cite{aldazabal}
 and  \cite{candelas3}  in section 3. 
\newline 
\section{The duality in K3 fibered Calabi-Yau 3-fold}
\par
~~~~~ We treat the extension from A-chain $n^0=2$ terminal case. 
$\CY_3=\CP^4(1,s,(s+1)(1,4,6))[12(s+1)]$ with 
$\K3=\CP^3(1,1,4,6)[12]$ fiber 
 have the same Hodge  numbers as those examples 
given  by  \cite{candelas2,yau}.
To see  why they coincide, we 
 compare    dual polyhedron of CY 3-folds. 
They  are not SL(4,{\bf Z}) equivalent.  
    We represent them  in the base where    
 one of   vertices  of 
 dual polyhedron  in the bottom  
    denotes the part of its weight 
for the hypersurface representation. It is
~$(-s,-s-1,-4(s+1),-6(s+1))$ in this case.
 The base manifold under the elliptic fibration is F$_2$  for $s=1$ case. 
F$_i$  denotes Hirzebruch surface.    
The dual polyhedron of 
F$_i$  has  three vertices~:~$ \{ \vec v_1,\vec v_2, 
-\vec v_1 -i\vec v_2 \}$.  
An example of integral points in the  dual polyhedron of F$_i$ is given by 
$\{\vec v_1=(1,0) ,(0,-1), \vec  v_2=(0,1),(0,0),(-1,-i)\}$.

  The Hodge numbers 
and the dualities in case (iii) are derived   
 by investigating the  extremal transition of the dual polyhedron 
of F$_0$ based CY3-fold in  \cite{candelas2}.
  
By using  dual polyhedra, we can find a 
 fibrations and base manifolds in some cases 
\cite{candelas1,candelas3,skarke,avram}.
 The upper suffix in $\nabla$  
 denotes the dimension of the lattice of a polyhedron.  
$\{(x_1,x_2,x_3,x_4)\} 
 ~\subset {}^4\nabla $~ forms the  
  dual polyhedron of CY3-fold.
$\{(x_1,x_2)\}~\subset  {}^2\nabla~ \subset {}^4 \nabla   $ represents  
the dual polyhedron of the base under the 
elliptic fibration of CY3-folds. 
In this paper, they are the blown up Hirzebruch surfaces.
$\{(x_2,x_3,x_4)\} \mid_{x_1=0}
 ~\subset {}^3\nabla ~ \subset {}^4\nabla $ 
 represents   
 the dual polyhedron 
of K3 fiber of CY3-fold.
$ \{ (x_3,x_4) \} 
\mid_{x_1=x_2=0} ~\subset {}^2\nabla $~$~ \subset {}^4\nabla~$ are 
 the dual polyhedron of common
 elliptic fiber of CY3-folds and K3fiber. However,
this is not         a sufficient condition of having elliptic fibration. 
The result of fibrations in CY3-folds  is given in table 1.
\begin{table}
\[
\begin{array}
{|l|l|l|l|l|l|}
\hline
\multicolumn{3}{|c|}
 {\rm CY3-folds ~of~  \cite{candelas2}} &
\multicolumn{3}{|c|}
  {\rm CY3-folds ~ of~ hyper ~ surface~ rep.}\\
\hline
\Delta n _T  & {\rm the~ base~ under~}T^2 & {\rm K3~ fiber} & s 
& {\rm the~ base~ under~}T^2 & {\rm K3~ fiber}
\\ \hline
0 &  {\rm F}_0  & \CP ^3 (1,1,4,6)[12]&1&{\rm  F} _2&CP ^3 (1,1,4,6)[12] \\ \hline
2 &  {\rm F} _2~ {\rm blown ~ up}    & \CP ^3 (1,1,4,6)[12]&2&{\rm  F} _2~
{\rm blown~up}&CP ^3 (1,1,4,6)[12] \\ \hline
3 &  {\rm F} _3 ~{\rm blown~ up}
 & \CP ^3 (1,2,6,9)[18]&3&{\rm  F} _2~
{\rm blown~ up}&\CP ^3 (1,1,4,6)[12]  \\ \hline
4 & {\rm F} _4~{\rm blown ~up}  & \CP ^3 (1,2,6,9)[18]&4& {\rm F} _2~ {\rm blown~ up}&\CP ^3 (1,1,4,6)[12]  \\ \hline
6 & {\rm F} _6~{\rm blown ~up} 
 & \CP ^3 (1,3,8,12)[24]&6& {\rm F} _2~{\rm blown ~up}&CP ^3 (1,1,4,6)[12] \\ \hline
8 & {\rm F} _8~{\rm blown ~up}
  & \CP ^3 (1,4,10,15)[30]&8&{\rm F }_2~{\rm blown~ up}&
\CP ^3 (1,1,4,6)[12]  \\ \hline
12 & {\rm F} _{12}~{\rm blown ~up} 
 & \CP^3(1,5,12,18)[36]&12&{\rm F} _2~{\rm blown~ up} & 
\CP ^3 (1,1,4,6)[12] \\ \hline
\end{array}
\]
\caption{ The kinds of fibrations of CY3-folds}
\normalmarginpar{N.B.~;~ The dual polyhedron of \CP$^3$(1,5,12,18)[36]
coincides with that of \CP$^3$(1,6,14,21)[42].}
\end{table}

In $s \geq2$ case, the additional integral points to F$_2$  are 
$\{(-i,-i-1) ~ {\rm with}~ 
\newline
2 \leq i \leq s,~(-1,-1)  \} $  within the base. 
Thus, for $s \geq2$ case, the base manifolds remain   blown up F$_2$ 
since ${ s+1 \over s} \leq 2$.
We can see  the Hodge number of the  base under the elliptic 
fibration and $n_T$ from dual polyhedron,   
$n_T+1 = $ $h^{1,1}({\rm F}_2~{\rm  blown ~up})=
d_1-2d_0=s+2 ~(s \geq 2),$
where $d_i$   is the number of 
i-dimensional cones  \cite{fulton}. 
$n_T$ coincides with case (iii) \cite{candelas2}.
On the other side,
K3 fibrations of case (iii) 
 change so that the base manifolds under elliptic fibrations  
 also   alter to F$_i$ blown up $(i \geq 2)$ with  
including  F$_0$  and F$_2$.
 In any case, the  bases  under  elliptic fibrations in case (iv)  are 
birationally equivalent to those in case (iii), 
which leads to the following identification.  
 (iii) and (iv) are connected by the extremal   
transitions  within  each sequences. $ \Delta n_T=0$ cases are   
deformed to each other 
by   the change of base manifolds, i.e.,  
F$_0 \leftrightarrow$ F$_2$ by using 
a non-polynomial freedom of deformation \cite{candelas2,gross}. 
Both CY 3-folds with $\Delta n_T=0 $ are the same manifold 
with double K3 fibrations \cite{gross}.

\begin{eqnarray}
\hat \nabla_{s=12, \Delta n_T=12} 
\supset 
\cdots  
\hat \nabla_{s=3,\Delta n_T=3} 
\supset 
\hat \nabla_{s=2,\Delta n_T=2} 
\supset 
&\hat \nabla_{s=1,\Delta n_T=0}&
\nonumber \\
&\updownarrow &
\nonumber \\ 
\nabla_{\Delta n_T=12} 
\supset 
\cdots
\supset
\nabla_{\Delta n_T=3}
\supset
\nabla_{\Delta n_T=2}  
\supset  
&\nabla_{\Delta n_T=0}&.  
\end{eqnarray}
 \noindent 
For $\Delta n_T > 0$, 
 blown up F$_2$  based CY3-folds
in case (iv) 
   also can be deformed to
 those of case (iii)  in \cite{candelas2}   by using a non-polynomial
freedom of deformation. 
$
\hat \nabla_{s=i, \Delta n_T =i} \leftrightarrow 
 \nabla_{\Delta n_T=i},
$
\noindent 
where 
 $\hat \nabla$ denotes  dual polyhedron of 
 case (iv) and $\nabla$ 
denotes those   of case (iii) 
\cite{candelas2}.   
The CY 3-folds 
 constructed by 
\cite{candelas2} and those with $s$ 
 are the same manifolds 
with  double K3 fibrations \footnote{They are represented by the 
same dual polyhedra in \cite{candelas3}.}. 
Each CY 3-fold  with   $\Delta n_T=i$ can be 
   represented  as a K3 fibration in two inequivalent ways as table 1.


The CY3-fold  with  $s=1$    is  
the terminal case of A-chains with $n^0=2$ in 
the  duality web \cite{aldazabal}. 
If duality exists, then 
their Hodge numbers  must 
satisfy the following conditions which 
comes from D=6  and  D=4  theories  as  the  
anomaly cancellation \cite{vafa1,candelas1,candelas2}.

\begin{eqnarray}
h_{2,1}\mid _{ G_2~ {\rm  in~ (iv)}  }&=&h_{2,1}^0 
\mid_{G_2^0=I~{\rm  in ~(i)}}
 -( a-b\Delta n_T) + {\rm dim.} G_2-29 \Delta n_T,
\\ \nonumber
h_{1,1} \mid_ {G_2 ~{\rm in~(iv)}}  &= & 
h_{1,1}^0 \mid_{G_2^0=I~{\rm in ~(i)}} + {\rm rank}G_2 
 + \Delta n_T,  
\\ \nonumber 
h_{2,1}^0 
\mid_{G_2^0=I~{\rm  in ~(i)}} &=&243,
~h_{1,1}^0 \mid_{G_2^0=I~{\rm ~in~ (i)}}=3.  
\end {eqnarray}
\noindent
where $G_1^0=G_1 =I$ up to U(1) in these cases.

\noindent 
This duality  between heterotic string and Type IIA string is 
 summarized in the 
 table 4, which  coincides with case (iii)  
\cite{candelas2}\footnote{  This  may imply  heterotic -
heterotic string duality.}. 

For CY3-folds side, $G_1$ and $G_2$ symmetries  
are  due to the   quotient singularities 
 of K3 fiber in the first column of table 1 
 rather than the quotient singularities of CY 3-folds  with s in   
table 2   
\footnote{
We compared  some  superpotentials of type IIB side in  case (iii) and 
those in case (iv) and examined  the possibility to derive the  
terminal  gauge symmetry   by the method of \cite{lerche}. 
We will give dual polyhedra of case (iv)  in the 
 next  paper to report about superpotentials
 more precisely.} .

\begin{table}[h]
\[
\begin{array}{|l|l|l|l|l|l|l|l|}
\hline 
s      & s=1   & s=2 & s=3 & s=4 & s=6   
& s=8 & s=12
\\ \hline 
{\rm quot. sing.}   & A_1              &A_1A_2      & A_1^2 A_3     
& A_1^4 A_4  & A_1^3 A_6  &  A_1^4 A_8 & A_1^6 A_{12}
\\ \hline 
\end{array}
\]
\caption{The quotient singularities in  K3=\CP$^3$(1,1,4,6)[12] fibered CY3-fold }
\end{table}

For K3$ \times$ T$^2$ side, 
they  are the subgroups of the E$_8 \times $ E$_8$  group 
for case (iii) \footnote{ 
They are perturbative gauge symmetries for case (iii) 
and may be  non-perturbative ones for case (iv) 
 as  $ \Delta n_T=0 $ case \cite{gross}.}.   
$a+ bn $ is calculated by the index theorem and denotes  
 the number of $G_1$ (or $G_2$) charged hyper multiplet fields, which is  
summarized in table 3~\cite{bershadsky,aldazabal,candelas1,candelas3}.
$n=n^0$ for $ G_1^0= \hat G_1$ case.
   The number of the charged hyper multiplets of $G_2$ 
  vanishes. We can see this by substituting  $n=-n^0$ for $G_2^0= \hat G_2$ case 
 and  $n=-\Delta n_T$  for $G_2=\hat G_2$  for A series ~\cite{candelas2,
candelas3}.

\begin{table}[h]
\[
\begin{array}
{|l|l|l|l|}\hline
G_1 & {\rm A~ series} &{\rm  B~ series} &{\rm  C~ series}  
\\ \hline
A_1 & 12n+32 &6k_1-39      & 4k_1-17  
\\ \hline
A_2 & 18n+54  & 10k_1-64    & 6k_1-26  
\\ \hline
A_3 & 22n+76 &12k_1-75      &  8k_1-33  
\\ \hline
A_4 & 25n+100 &14k_1-84      & 9k_1-35 
\\ \hline
D_4 & 24n+96 &                &
\\ \hline
D_5 & 26n+124 &15k_1-90      & 10k_1-39   
\\ \hline
E_6 & 27n+162 &16k_1-94      &     
\\ \hline
E_7 & 28n+224 & & \\ \hline
\end{array}
\]
\caption{  The review of the number of the charged hyper multiplets.} 
\end{table}

\begin{table}[h]
\[
\begin{array}{|l|l|l|l|l|l|l|l|l|l|}\hline 
s      & U(1)^4 \times G_2   & h^{1,1}& h^{1,2}& k_1& k_2  
& n_ T^0&\Delta n_T& n_T &n^0 \\ \hline 
s=1    & U(1)^4 \times (G_2^0=I)             &3      &243     & 12 & 12   &  1&0&1   &
2  \\ \hline
s=2    & U(1)^4 \times I             &5       &185     & 12 & 12-2&  1&2&3& \\ \hline
s=3    & U(1)^4 \times A_2   &8       &164     & 12 &12-3 &  1&3&4 &\\ \hline
s=4    & U(1)^4 \times D_4   &11      &155     & 12 & 12-4 &  1&4&5 &\\ \hline
s=6    & U(1)^4 \times E_6   &15      &147     & 12 & 12-6 &  1&6&7 &\\ \hline
s=8    & U(1)^4 \times E_7   &18      &144     & 12 & 12-8 &  1&8&9 &\\ \hline
s=12   & U(1)^4 \times E_8   &23      &143     & 12 & 12-12 &  1&12&13 &\\ \hline
\end{array}
\]
\caption{The duality of (iv) about 
K3=\CP$^3$(1,1,4,6)[12] fibered CY3-fold }
\end{table}

\section{Discussion and Conclusion}
\par In this article, we have studied the property of CY3-folds 
with $ s\geq 2$ case  from A series. 
They are the same CY 3-folds as those constructed in \cite{candelas2}, 
which have  double K3 fibrations. 
\newpage
For the  puzzle III,  the conclusion is that  
the hypersurface representations  with  
 $s \geq 2$ cases 
from  B series and C series in $n^0=2$    satisfy 
the duality  of terminal B and C series  with extra tensor multiplets 
\footnote{
For B and C sequences, the modification of eq. (6) is necessary. 
The change in the Hodge numbers, $h_{2,1}$ according to the change   
of the terminal group is not only the difference of 
dim. $G_2^0$~( or dim. $G_2$)~\cite{aldazabal,candelas3}. 
Thus we compare  the Hodge numbers of CY3-folds  which  have 
 the same terminal gauge symmetry.}.
 For example, B and C  series  in s=2, their 
Hodge numbers can be interpreted as those  with  shrinking of 
 two  instantons from   
 $n^0=2$    case   as the  ref. of \cite{candelas3}.  
$s \geq 3$    
cases  are also explained by shrinking $\Delta n_T$ instantons  
 from the terminal case with the   same  $\hat G_2$  symmetry.

\begin{eqnarray}
&\Delta h_{2,1}&= -h_{2,1}^0 \mid_{G_2^0=\hat G_2~{ \rm terminal~in~(i)}}
+h_{2,1} \mid_{G_2=\hat G_2~{ \rm terminal~in~(iv)}},
 \nonumber \\
&\Delta h_{1,1}&= -h_{1,1}^0 \mid_{G_2^0=\hat G_2~{ \rm terminal~in~(i)}}
+h_{1,1} \mid_{G_2=\hat G_2~{ \rm terminal~in~(iv)}},
 \nonumber \\
&{\rm B~ series}&~ : -\Delta h_{2,1}=17 \Delta h_{1,1},
 ~{\rm C~ series} : -\Delta h_{2,1}=11 \Delta h_{1,1}.
\end{eqnarray}
\noindent 
This situation is quite similar  to  that of A series.  
They are interpreted as the dualities obtained by
 unhiggsing of $U(1)$ and $U(1)^2$ 
from (iv) in A series.
 
For puzzle IV, 
Candelas, Perevalov and Rajesh  seemed to
construct B and C versions of (iii)   and derived 
the Hodge numbers already, though they do not write them  
explicitly \cite{candelas3}.
 According to the  description,
  Hodge numbers of them  are the same as table 5 and table 7.


  \begin{table}
  \[
  \begin{array}{|l|l|l|l|l|l|l|l|l|l|}\hline
  s & U(1)^4  \times U(1)  \times G_2 & h^{1,1}&h^{1,2} &
  k_1&k_2&{n_T}^0&\Delta n_T& n_T &n^0\\
  \hline
  s=1    & U(1)^4 \times U(1)   \times (G_2^0=I)           &   4    & 148    & 9 & 9 &1   
  &  0  & 1& 2\\ \hline
  s=2 & U(1)^4  \times U(1) \times I                  & 6       & 114      
  & 9 & 9-2    
  & 1&2&3 & \\ \hline
  s=3&U(1)^4   \times U(1)\times A_2 & 9 & 101 & 9 & 9-3  
  & 1&3&4 &\\ \hline
  s=4&U(1)^4   \times U(1) \times D_4 & 12 & 96 
  & 9    & 9 -4 
   & 1&4&5&  \\ \hline
  s=6&U(1)^4   \times U(1)\times E_6  & 16 & 92 
  & 9    & 9 -6 
   & 1&6&7&  \\ \hline
  \end{array}
  \]
  \caption{ The duality of (iv) about K3=\CP$^3$(1,1,2,4)[10] 
  fiberd CY3-fold  }
  \end{table}

  \begin{table}
  \[
  \begin{array}{|l|l|l|l|l|l|l|l|l|}\hline
  n^0 & U(1)^4  \times U(1)  \times G_2^0 & h^{1,1}&h^{1,2} &
  k_1&k_2&{n_T}^0&\Delta n_T& n_T \\
  \hline
  n^0=0&U(1)^4\times U(1) \times I                      & 4       & 148      &9    & 9 &1 
  & 0&1  \\ \hline
  n^0=3&U(1)^4   \times U(1) \times A_2 & 6 & 152 & 9+3 & 9-3 
  & 1&0&1 \\ \hline
  n^0=4&U(1)^4  \times U(1)\times D_4   & 8 & 164 
  & 9+4    & 9 -4 
   & 1&0 &1  \\ \hline
  n^0=6&U(1)^4   \times U(1) \times E_6 & 10 & 194 
  & 9+6    & 9 -6 
   & 1&0&1  \\ \hline
  \end{array}
  \]
  \caption{ The duality of (i) about  terminal CY3-folds in B series }
  \end{table}

  For the puzzle V,  we examine K3 fiberd CY3-folds 
  whose s=1 case is  
  the terminal case   in the A-chains with $n^0=4$. 
  Their K3 fiber is 
  \newline
   K3=\CP$^3$(1,2,6,9)[18].
  The dual polyhedron of $s=2$ case  has  
    following additional points  
  $(-2, -6,-18, -27)$, $(-1,-3,-10,-15)$ and $(-1, -3,-9,-14)$, which  imply  
  $\Delta n_T$=1. $ h_{1,1} \mid _{s=2}-h_{1,1}^0 \mid _{s=1}=3$.
   The terminal group should change 
  from $D_4$ to $G_2$ with rank 6. We can not find 
  $G_2$  which satisfies the extension of eq. (6).
   It seems that it needs another idea to see the 
  duality in this case. 
  \newline
  {\bf Acknowledgement}
  ~~~~~We gratefully acknowledge fruitful conversations with  
  M. Kobayashi, S. Hosono and N. Sakai. We would like to thank 
 K. Mohri mostly for useful discussions.

\begin{table}
\[
\begin{array}
{|l|l|l|l|l|l|l|l|l|l|} \hline
s& U(1) \times U(1)^2 
\times G_2 & h^{1,1}& h^{1,2}&k_1&k_2&{ n_T }^0& \Delta n_T & n_T
&n^0
\\ \hline
s=1 & U(1)^4\times U(1)^2 \times (G_2^0=I) &5 & 101&7&7&1&0 &1&2\\ \hline
s=2 & U(1)^4 \times U(1)^2 \times I & 7& 79& 7& 7-2 &1&2&3 & \\ \hline
s=3 & U(1)^4  \times U(1)^2 \times A_2 & 10 &70& 7 & 7-3 &1&3&4 & \\ \hline
s=4 & U(1)^4  \times U(1)^2 \times D_5 & 13 &67& 7 & 7-4 &1&4&5 & \\ \hline
s=5 & U(1)^4  \times U(1)^2\times E_6  & 17 &65& 7 & 0 &1&5&6 & \\ \hline
s=6 & U(1)^4  \times U(1)^2\times E_6  & 17 &65& 7 & 0 &1&6&7 & \\ \hline
\end{array}
\]
\caption{The duality of (iv) about K3=\CP$^3$(1,1,2,2)[6] fiberd  CY3-fold 
}
\end{table}

\begin{table}
\[
\begin{array}
{|l|l|l|l|l|l|l|l|l|} \hline
n^0& U(1) \times U(1)^2 
\times G_2^0 & h^{1,1}& h^{1,2}&k_1&k_2&{ n_T }^0& \Delta n_T & n_T

\\ \hline
n^0=0 & U(1)^4\times U(1)^2\times I &5 & 101&7&7&1&0 &1\\ \hline
n^0=3 & U(1)^4  \times U(1)^2\times A_2  & 7 &103& 7+3& 7-3 &1&0&1 \\ \hline
n^0=4 & U(1)^4  \times U(1)^2 \times D_5 & 10 &110& 7+4 & 7-4 &1&0&1 \\ \hline
n^0=5 & U(1)^4  \times U(1)^2 \times E_6 & 12 &120& 7+5 & 0 &1&0&1 \\ \hline
n^0=6 & U(1)^4  \times U(1)^2 \times E_6  & 11 &131& 7+6 & 0 &1&0&1 \\ \hline
\end{array}
\]
\caption{The duality of (i) about  terminal CY3-folds in C series  
}
\end{table}



\newpage

\begin{table}[h]
\begin{center}
\begin{tabular}{|l| c c c |} \hline
$G_2^0~ \backslash~ G_1^0 $ & $ G_1^0$ & $ \stackrel {\rm (ii)}
 {\longrightarrow}$ &  $ G_1^0=\hat G_1 $ \\ \hline 
$G_2^0=I$  &  $ I \times I$ & $\cdots$ & $\hat G_1 \times I$ \\  
$ ( n^0=0, 1,2 )$ & $\cdot$ & $ \cdots$ &  $\cdot$ \\
& $\cdot$ & $\cdots$ & $\cdot$ \\
$\big \downarrow $ (i)  &  $ \cdot$ & $\cdots$  & $\cdot$  \\ 
& $\cdot$ & $\cdots$ & $\cdot$ \\
$G_2^0= \hat G_2 $  & $\cdot$ &$ \cdots$ & $ \cdot$  \\
($ 12 \geq  n^0 \geq 3 $ ) & $ I \times  \hat G_2 $ & $\cdots $ & 
$\hat G_1 \times \hat G_2 $  \\ \hline  
\end{tabular}
\end{center}
\end{table}
Fig. 1~ The gauge symmetries from (i) and (ii)

\begin{table}[h]
\begin{center}
\begin{tabular}{|l| c  |} \hline

$G_2^0~ {\rm or  }~G_2~ \backslash~ G_1^0~ {\rm or}~G_1 $ & $G_1^0=G_1=I $  \\ \hline 
$G_2^0=I$  &  $ G_1^0 \times G_2^0= I \times I $ \\  
$ ( n^0=0,~\Delta n_ T =0 )$ & $\cdot$  \\
& $\cdot$  \\
$ \big \downarrow $ (iii)  &  $ \cdot$ \\ 
& $\cdot$  \\
$G_2=\hat G_2 $  & $\cdot$   \\
($  \Delta n_ T  \geq 3 $ ) & $G_1 \times  G_2 =I \times \hat G_2 $  \\ \hline  
\end{tabular}
\end{center}
\end{table}
Fig. 2~ The gauge symmetries from (iii)

\end{document}